# Spatiotemporal Analysis of Meteorological Drought Variability in the Indian Region Using Standardized Precipitation Index

M. Naresh Kumar[1*], C.S. Murthy[2], M.V.R. Sesha Sai[2] and P.S. Roy[2]

[1] Software Development & Database Systems Group

[2] Remote Sensing & GIS Applications Area

National Remote Sensing Centre, Hyderabad 500 625, India

**Abstract**

Grid (1° latitude x 1° longitude) level daily rainfall data over India from June to September for the years 1951 to 2007, generated by India Meteorological Department, was analyzed to build monthly time series of Standardized Precipitation Index (SPI). Analysis of SPI was done to study the spatial and temporal patterns of drought occurrence in the country. Geographic spread of SPI derived Area under Dryness (AUD) in different years revealed the uniqueness of 2002 drought with wide spread dryness in July. Mann-Kendal trend analysis and moving average based trends performed on AUD indicated increasing trend in July. The area under moderate drought frequency has increased in the most recent decade. ―Ranking of years based on Drought Persistency Score (DPS) indicated that the year 1987 was the severe-most drought year in the country. The results of the study have revealed various aspects of drought climatology in India. A similar analysis with the SPI of finer spatial resolution and relating it to crop production would be useful in quantifying the impact of drought in economic terms.

[*] Correspondence to: M. Naresh Kumar, Senior Scientist, Software and Database Systems Group, National Remote Sensing Centre, Hyderabad, Andhra Pradesh, India 500 625. E-mail: nareshkumar_m@nrsc.gov.in

1. Introduction

South west summer monsoon, spreading from June to September is a grand period of rainfall in India as monsoonal torrents supply about 80% of India's annual rainfall (Chang, 1967; Bagla, 2006). The monsoon reaches south India, generally by the end of May or first week of June and progresses to northern parts by the end of June or first week of July. The withdrawal of monsoon starts in the end of September. Rainfall during this season plays a vital role in economic development, disaster management and hydrological planning in the country.

Time series rainfall data analysis of a region helps in better understanding its drought climatology. In addition to detection of changes in intensity, magnitude and duration of droughts, identification of frequently drought affected regions also plays an important role in drought management. Further, past performances provide indications on the future scenarios. Information on spatial and temporal dimensions of drought occurrence and its spread enables designing of more focused management tasks. Therefore, systematic understanding of drought climatology is indispensable for evolving efficient drought management strategies particularly in tropical regions like India. A number of studies have been reported on analysis of time series rainfall data for characterizing the drought events in terms of periodicity and magnitude. Such long term rainfall data analysis would also result in the assessment of drought vulnerability. The dominant rainfall patterns for entire India were evolved through time series rainfall data from 1871-1990 analysis, using map-to-map correlation, fuzzy *c*-means clustering and empirical orthogonal functions (Kulkarni and Kriplani 1998, and Kulkarni *et al*., 1992). It was observed that the rainfall from monsoons over entire India has no relation with its spatial distribution.

There are many indices to measure meteorological dryness, such as simple rainfall deviation from historic normal, Palmer Drought Severity Index (PDSI) and Standardized Precipitation Index (SPI). Among these indices, SPI is being widely used in recent years because of its computation simplicity and interpretation fidelity. For representing dryness a simple measure is better than using complex hydrological indices (Oladipio, 1985). SPI is a simple and more effective method for studying drought climatology (Lloyd-Hughes and Saunders, 2002). Both SPI and PDSI indicated temporal changes in proportion of area experiencing drought in Europe (Lloyd-Hughes and Saunders, 2002). Pai *et al.* (2010) evaluated district wise drought climatology in India using SPI. Briffa *et al.* (1994) characterized drought across Europe using PDSI derived from time series rainfall data from 1892-1991 at a spatial resolution of 0.5°. Analysis of time series PDSI in Hungary revealed occurrences of more droughts in the end of time series (Szinell *et al*., 2002). Bordi *et al*. (2001) computed SPI with time series rainfall data from 1948 to 1981 and analyzed regional drought patterns in Mediterranean. Rodriguez-Pubela *et al*. (1998) analyzed annual precipitation observations through empirical orthogonal functions and derived four regional precipitation regimes. Sirdas and Sen (2003) used kriging technique to generate spatial maps of rainfall and SPI to characterize drought intensity and magnitude in Trakya region, Turkey. Vicente-Serrano *et al*. (2004) analyzed drought patterns in Spain using SPI and observed significant increase in the area under drought from mid to northern area. The spatial patterns of drought were found to be very complex in rest of the areas. Iran's precipitation climate was regionalized based on rainfall data analysis using factor analysis and clustering (Dinpashoh *et al*., 2004).

The objective of the current study is to build a long term monthly SPI time series from the rainfall data and to analyze the SPI derived Area under Dryness (AUD) for drought spread, drought frequency and drought persistence. The emphasis is on highlighting the spatial and temporal variations in the occurrence of meteorological dryness. Although, several studies were reported in India, on drought analysis based on rainfall data (Chowdhury *et al*., 1989; Sen *et al*., 1997; Guhathakurta, 2003; Sinha and Shwale, 2001; Pai *et al.,* 2010), the present study finds relevance due to its larger coverage, long term data analysis and adoption of robust indicator like SPI. Further, since the south west monsoon is a key determinant of the performance of kharif agricultural season, the results of the study become more relevant and significant, particularly for drought management. Section 2 of the paper describes methodology for computing SPI using regional probability functions, and trend analysis using Mann-Kendall trend coefficient. In section 3, results and discussion are presented covering the analysis pertaining to drought spread, inter-annual variability of AUD, trends in AUD, spatial drought frequency and drought persistence. Conclusions of the study are presented in section 4.

## 2. Methodology

Daily rainfall data from June to September for 57 years (from 1951 to 2007) generated by India Meteorological Department ( IMD) at a grid size of 1° latitude x 1° longitude was the input data in this study. The details of gridded rainfall data generation are available in Rajeevan *et al.* (2006). Monthly Standardized Precipitation Index (SPI) computed using Pearson III distribution for different years.

There are three important steps in the computation of SPI; (1) development of regional probability functions based on regional rainfall patterns i.e., regionalization approach particularly while using the data over larger geographic areas representing diverse rainfall patterns, (2) selection of suitable probability distribution and (3) estimation of parameters of the selected distribution. We adopted regionalization approach with Pearson III distribution by deriving the distribution parameters through L-moments method following Guttman *et al*., (1993) and Guttman, (1999).

Regional probability functions are developed after identifying homogeneous rainfall regions. For example, National Climatic Data Centre (NCDC) prepared National Drought Atlas for the U.S. Department of Energy using monthly precipitation and temperature data for 1219 stations by adopting hierarchical clustering techniques (www.iwr.usace.army.mil/iwr/atlas). IMD categorized the country in to 36 meteorological sub-divisions based on rainfall patterns. Parthasarathy *et al*. (1995) grouped India into five homogeneous regions based on spatial contiguity, actual seasonal rainfall, global circulation parameters etc. Since the generation of rainfall homogeneous zones is not the main task of the current study, five broad zones based on the average rainfall are considered; Region 1 with <500 mm of rainfall, Region 2; 500-750 mm, Region 3; 750-1000 mm, Region 4; 1000-1500 mm and Region 5; >1500 mm of rainfall were considered for analysis, with due consideration to the spatial contiguity of these regions.. The rainfall intervals of these regions are generally in agreement with the standard rainfall variability in India. According to the reports of the Ministry of Agriculture, Government of India, the low rainfall region receiving less than 750 mm of rainfall represents 33% of cropped area; rainfall of 750-1125 mm represents 35% of cropped area. Similarly, rainfall of 1125-2000 mm represents 24% of cropped area and >2000mm represents 8% of sown area. The purpose of rainfall zoning in this study is to develop regional probability functions to evolve a robust SPI.

## 2.1 Estimation of the parameters of Pearson III Distribution using L-Moments

L-Moments are the estimates of moments using linear order statistics, for relating a probability distribution to the observation of the physical process. The L-moment estimates are more robust as compared to the conventional moments in the presence of outliers (Royston, 1992; Sankarasubramanian and Srinivasan, 1999; Ulrych *et al.*, 2000). The L-moments are less sensitive to the effects of sampling variability. The L-moments are used to characterize a wider range of distributions than the conventional moments. Practically they are less subject to bias in estimation, and they approximate their asymptotic normal distribution more closely. The parameters estimated through L-moments are more accurate than the maximum likelihood and least square estimates. This property of L-moments has been extensively used in the regionalization processes (Guttman, 1993; Hosking and Walis, 1989; Andamowski, 2000).

The L-moments of the regions were computed using a MATLAB subroutine. The parameters of the Pearson III distribution were computed using FORTRAN subroutines of Hosking (1990) invoked through an interface developed in MATLAB.

## 2.2 SPI based drought classes

Ranges of SPI values and the corresponding drought intensity levels proposed by Mc Kee *et al*. (1993) were adopted in this study. Two drought classes namely, moderate drought – with SPI ranging from -1 to -1.5 and severe-extreme class with SPI <-1.5 were used to represent dryness. Summation of moderate drought and severe-extreme drought classes was done to represent the total AUD. That means, all the grid cells having SPI value less than or equal to -1.0, in a given month, constitute the AUD.

## 2.3 Methods for Smoothing Time Series Data and Trend Detection

The SPI derived AUD time series data was first analyzed for presence of autocorrelations. Durbin-Watson d test was performed for each month separately. The computed $d$ value and the corresponding $d_L$ and $d_U$ were applied to the decision rules (Gujarati, 1988) and found that the autocorrelation were not statistically significant in any month. Mann-Kendall trend test based on Sen's Method (Sen, 1968) was selected as it was well suited for analyzing trends in data over time (Gilbert, 1987; Gregory, 2004; Picarreta *et al.*, 2004) and it does not require any assumptions as to the statistical distribution of the data (e.g. normal, lognormal, etc.). To remove noise and bring out significant trends in the time series data, local weighted scatter plot smoothing (LOWESS) a non-parametric technique based on local polynomial fits ( Cleveland, 1979) was implemented before subjecting the data to Mann-Kendall Trend test. LOWESS is a widely used smoothing technique for detecting trends (Wang *et al*., 2003; Humphrey *et al*., 1999; Abaurrea *et al*., 2001) in the time series data. A smoothness parameter $f = 0.06$ (Boaz Kaunda, 2003) was selected as it was found to adequately smoothen the data without distorting the main temporal patterns.

Mann-Kendall Statistic *S* was computed after arranging the data in rank order and correcting for ties. Significance of trend was checked at 95 percentile to reject the null hypothesis H0: There is no trend in the different dryness classes in the region. The trend is increasing (*I*) if the *S*>0 and confidence in trend is greater than 95%. The trend is decreasing (*D*) if *S*<0 and confidence in trend is greater than 95%.

3. **Results and Discussion**

The results of the study are discussed under four sub-topics; (1) drought spread in the time series with emphasis on geographic spread of drought in major drought years (2) inter-annual variability of AUD, (3) regional trends in the AUD, (4) Trend reversals in AUD time series (5) drought frequency and (6) drought persistence.

**3.1 Drought spread in different years**

The main agricultural season namely kharif season extends from June to October/November and depends largely on the rainfall from south west monsoon. July month represents the peak sowing period, August represents the active growing period of crops and September represents the maximum vegetative phase or early reproductive phase of rainfed crops. Harvesting of rainfed crops is mostly done in the month of October or November in different parts of the country. Keeping in view, the crop phenology and crop calendar, rainfall during July, August and September months plays a vital role in dry land agriculture. Rainfall in the month of June is essential in triggering sowings in the country. AUD during July to September months has direct bearing on the agricultural situation in the country. Although sowing starts in the month of June, larger proportion of the agricultural area is sown during July month. Therefore, rainfall during July month is very critical for timely completion of crop sowings as well as for the sustenance of early sown crops. Similarly, dryness in August and September months strongly influence crop growth and performance. Therefore, drought spread analysis was carried out for each month separately.

The spatial extent of dryness in different months in some of the prominent drought years of the country was depicted in Figure 1. During June (figure1a), the AUD was mostly confined to central and northern parts of the country. Also, the dryness in June occurred mostly in the earlier years of the time series. In recent years the geographic spread of dryness in the month was to lesser extent. In July month (figure 1b), the year 2002 stands unique with a large number of dry grids covering many parts of the country, compared to any other year in the time series. The drought spread in 1987 and 2006 showed contrasting patterns with a symmetrical shift in the dry grids in vertical mode i.e., east and west side. During August (figure 1c), the dryness of 1968 was concentrated in the southern half and in the western parts of the country. In 1979, dryness was located mostly in the northern part of the country. The most prominent drought year of 1987 has dryness located as distinct clusters in north, east and west sides. The September dryness (figure 1d) was mostly concentrated to the north western part of the country, in synchronous with the receding pattern of the south-west monsoon. Thus, the prominent drought years showed spatial uniqueness in the geographic spread of dryness.

The AUD categorized under (a) moderate (SPI -1 to -1.5) and (b) severe-extreme drought (SPI <-1.5) classes during 1951-2007 time period for the months of June, July, August and September separately are depicted in Figure 2a-2d. The significance of inter annual variability of AUD in all the four months is evident from visual examination of these Figures. The coefficient of variation (CV) of AUD which reflects the average rate of change over the time series indicated larger variations to the extent greater than 70% in July and September months. July month represents the most active phase of monsoon and September represents the receding phase of monsoon. From agriculture point of view, July is more critical because of its influence on crop sowings. Dryness in September if not preceded by dryness in the August has less bearing on the crops. Thus, the uncertainty in AUD vis-à-vis rainfall during July month is more when compared to other months. The CV was found to be lesser during the months of June (which is the beginning of the season) and August (which is the active crop growing phase), when compared to other months. In absolute terms, the CV values of June and August were at slightly less than 50%, which signifies considerable variability and uncertainty in the occurrence of dryness. The dryness in the month of June affects the crop sowings to some extent, particularly the early sown crops. The dryness and its variability in the month of August which represents the active crop growing period has more deleterious effect on the performance of crops.

The year wise comparison of June AUD shows the years 1965, 1969, 1962, 1992, 1968, 1987 and 1974, with more dryness(>100) grids in each year. Similarly, the lower AUD was observed in the years 1971, 2001, 1970, 1978, 1980, 1956, 1993, 1977, 1961. The AUD in these years were less than 30 grids in each year. The AUD in June was widespread in the decade of 1960s, whereas it is interesting to note that in the recent decade comprising of the years 1997-2007, it was significantly less indicating reduced dryness, in the month.

The year 2002 has more number of grids under dryness in July month compared to rest of the years in the time series, followed by 1987. Even though 1987 was the second-most widespread drought year, the number of dry grids in the year are approximately half that of year 2002. Further, the proportion of area under severe-extreme dryness was very large in 2002. Therefore, July dryness during 2002 is geographically wide spread and more intensive and hence remains as an unprecedented drought event. The drought of 2002 in India is unique due to acute rainfall deficiency in 18 out of 26 states, as a result of which crops could not be sown (www.imd.gov.in, www.agricoop.nic.in). The normal practice of declaring drought at the end of the season in October after observing the performance of crops was not practiced in 2002, as most of the states adopted the unusual approach of declaring drought by the end of July or first week of August (Samra, 2004; Rathore, 2005).

The other years having extensive dry areas in July month include 1987, 2006, 1955, 2004, 1972 and 1979. Among these years - 1987, 1972 and 1979 were reported to be wide spread drought years in the country (DFID, 2008). In the other three years namely 1955, 2004 and 2006, dryness was confined to July alone and hence they could not be considered as drought years.

The AUD in the month of July was very low in the years 1964, 1960, 1988, 1961, 1976. In general, more area under July dryness was observed only during recent years. As compared to earlier decades, the number grids with dryness in July are higher in the recent decades.

Larger area under dryness was detected in the August month of 1993, 1968, 1979, 1972, 1987, 2005 and 2000. Of these years, in 1993, 1987, 1972 and 1968 the proportion of area under severe-extreme dryness was significant. The prominent drought years in the country are 1987, 1972, 1979 and 1968 (www.agricoop.nic.in).

The AUD in September month was mostly under moderate category. The years with more area include 1952, 2001, 1968, 1987, 1986, 1974 etc and the years with very less area under dryness include 2006, 1955, 1975, 1983 etc. The years with more dry areas such as 1987, 1974 and 1968 happened to be the drought years in the country.

**3.2 Trends in AUD**

Area under (a) moderate dryness, (b) severe-extreme dryness and (c) total dryness derived from SPI thresholds were subjected to LOWESS algorithm for smoothening. Mann-Kendal trend coefficient was computed for each of the four months.

Decreasing trend in the total area under dryness (moderate + severe-extreme classes) was evident in June month, with severe-extreme dryness showing no trend and moderate dryness showing decreasing trend. That means, rainfall tends to be closer to normal in the initial month of monsoon. The areas under all the three classes of dryness - moderate, severe-extreme and total, showed increasing trends in July. Therefore, increasing uncertainty in the rainfall of July, a critical month for agricultural season was evident. During August month, the AUD reveals no significant trend, although the area under severe-extreme dryness shows increasing trend. Since the area under this class is very less, this does not signify serious implications. There was no significant trend in any dryness category in September month. Significant inter annual fluctuations in the withdrawal pattern of monsoon in the month could cause no significant trend in the dryness.

Thus, to summarize, the area under dryness was decreasing during June and increasing during July followed by no significant trend in the other two months. Mooley and Parthasarathi (1984) found that at all India rainfall was without any trend and random in nature. Trends in rainfall at different spatial scales were also observed by Parthasarathy (1984) and Rupa Kumar *et al*. (1992).

Change point analysis was performed to detect the differential trends in the AUD within the time series. Significant trend reversals were detected in 3 cases; (1) area under severe-extreme dryness in June showing no trend in the total time series, increasing trend during 1951-1965 and decreasing trend during 1992-2007, (2) area under severe-extreme dryness in September showing no significant trend in the overall time period but increasing trend during 1954-1986 and (3) area under total dryness (moderate + severe-extreme) showing no significant trend in the overall time period but increasing trend during 1954-1986.

### 3.2.1 Trends with moving averages

Trend analysis of time series AUD was also performed with moving average technique. 5-year and 7-year moving averages were computed and trend lines were fitted for the AUD of each month separately. The details of trend equations were presented in Table 1. Among the trend lines of four months, only that of July was statistically significant. The trend line of July showed increasing trend in the AUD. The trend coefficient of June is unique since it assumed negative value indicating decreasing trend in the AUD in contrast to rest of the months. The trend lines of August and September months were almost flat with very small increases from year to year as indicated by smaller trend coefficients.

### 3.3 Drought frequency

Drought frequency, in this study, was measured by the number of years, a grid cell experienced dryness (SPI <-1.0) in the total time series of 57 years. Grid cells which experienced dryness in 19-29 years out of 57 years i.e., drought frequency of 2-3 years and the grids under dryness in 10-14 years out of 57years i.e., drought frequency of 4-5 were identified and mapped. Month wise drought frequency maps thus developed reveal significant number of grids belonging to high frequency (2-3 years) in the months of June and September (Figure 4a). Such grids were mostly distributed in central India. During July and August months such high frequency grids were less in number showing isolated distribution.

Moderate drought frequency (drought occurring once in 4-5 years), was observed in a significant number of grid cells, particularly in June and July months (Figure 4b). In June, such grids were mostly distributed in the Central and the North India which could be due to fluctuations in the arrival time of monsoon and variations in the quantity of rainfall. Moderate drought frequency grids in July month were more in number compared to other three months and were well distributed throughout the country. This shows the growing uncertainty in July rainfall. During August month, the drought frequency grids were mostly located in the southern and northern India. In September month, such frequent drought grids were mostly located in the north and central India.

Decade wise analysis of drought frequency was performed to bring out the variations within the time series. The interval of decade is selected for convenience and easy perception of interpreted results. Kiem and Frank, 2004 and Gregory *et al*., 2004 have identified changes in the climate occurring on decadal to multi-decadal time-scales. The decadal analysis facilitates identification of decades with significant changes. The time series data in this study was divided in to five decades – decade ending 1967, 1977, 1987, 1997 and 2007.

In each decade, the grid cells under dryness i.e., SPI of less than -1.0, were identified. Drought frequency indicating the number of years under dryness was represented in three categories namely, (1) Low frequency with less than two years of dryness in 10 years, (2) Moderate frequency with 3-4 years of dryness and (3) High frequency with more than 4 years experiencing dryness in a decade indicating drought in almost every year or alternate year. The area under each frequency class in terms of number of grid cells was computed for each decade. The decade wise areas under three drought frequency classes for June, July, August and September months were shown (Figures 5a-5d).

In June, low frequency area was significantly higher in the decade ending 2007 compared to all other decades. In all the other decades such area was more or less equal. Area under moderate frequency and high frequency categories were significantly less in the decade of 2007 compared to the rest. In the remaining 4 decades, the area with moderate and high frequency droughts do not show significant differences. Thus, the area with low frequency dryness was more and the area with high frequency dryness was less during June in the recent decade. This unique feature of recent decade compared to rest of the time series indicates less varying rainfall in June in recent years.

During July, the area under low frequency drought was decreasing from 1960s with a larger reduction during the recent decade of 1996-2007. Similarly, area under moderate drought frequency was on constant rise from the past with significant increase in the recent decade. There was slight increase in the area with high frequency drought. There was significant increase in the area under moderate frequency drought in the month.

The area under different categories of drought frequencies has shown insignificant variations during August month in different decades. Most of the area in different decades was under low frequency drought. None of the five decades exhibited any uniqueness or discernable pattern in any drought frequency class.

Drought frequency during September month revealed mixed trends. Area with less frequent drought was higher in the decades ending 1967 and 1997. In the remaining three decades it was slightly less. The decade ending 1977, was characterized by more area under high frequency drought.

### 3.4 Drought Persistence

The magnitude of drought is determined by geographic spread of dryness as well as persistency of drought conditions in the time domain. Continued dryness over longer periods of time is called persistent drought. Unsustained drought conditions i.e., dryness that exist for shorter time periods are called non-persistent or intermittent droughts. While assessing the magnitude of drought, one has to account for the area under persisting and non-persisting drought. Persistent dryness results in more deleterious effect on agriculture, surface water and ground water. Therefore, drought persistence assumes greater importance in the drought management.

During the four months of south west monsoon season (June to September), the drought conditions occur and persist in different ways on a monthly time scale; (1) 4-months drought persisting over all the four months, (2) 3- months drought occurring in two possible combinations, June-August and July-September (3) 2-months drought occurring in 3 possible ways, June+July, July+August, August+September and (4) 1-month drought, i.e., non persistent drought occurring in any one of the four months. In terms of impact, the 4-months drought has maximum impact whereas single month or intermittent droughts have relatively lesser impact.

Drought Persistency Score (DPS) was developed in this study, by assigning linear weights to the areas under each persistency class. The weighing factor for 4-month drought is 4, for 3-month drought are 3 and so on. Thus, DPS is a direct indicator of magnitude and intensity of drought for the season as a whole. In each year, the number of grids under different combinations of persistent drought and non-persistent dryness during June to September were identified and DPS was computed.

The DPS for different years arranged in ascending order was presented in Figure 6. The highest values of DPS in 1987, 1974, 1972 and 2002 indicate more intensive drought in these years. In these four years, 1987 monsoon season stands unique with very high values of DPS. Another uniqueness of 1987 was its very high proportion of persistent drought score in the total score, clearly signifying the highest magnitude of drought conditions. The drought of 1987 was one of the worst droughts of India, with a rainfall deficiency of 19 per cent (DFID, 2008 and www.agricoop.nic.in). The drought has affected 60% of crop area. As per the reports of the Ministry of Agriculture, monsoon rainfall was less than normal by 24 % in 1972, 12% in 1974 and 19% in 2002. In these years also, the proportion of persistent drought score was considerably high reflecting the drought intensity. The 2002 drought ranks fifth in terms of magnitude in the country (DFID, 2008). The second set of years with higher values of DPS include; 1968, 1979 and 1991. The rainfall deficiency in 1979 drought was 19% which has resulted in the reduction in the production of food grains by 19% (DFID, 2008 and www.agricoop.nic.in). The lowest values of DPS could be observed in 1961, 1983, 1977, 1964, 1975 and 1981 which indicates very less magnitude of compared to rest of the years in the time series. In these set of years, the proportion of persistent score in the total score was also less, signifying very low magnitude of drought. On the basis of DPS, the time series 1951 to 2007 can be grouped in to different classes as shown in Table 2. It is evident that there were many years with moderate magnitude in the recent decade. Most of the years in 1950s and 1960s recorded lower DPS score indicating lower magnitude.

4. Conclusion

Regionalized Pearson III distribution with L-moments based parameter estimation was adopted to compute SPI from the all India 1° latitude x 1° longitude gridded rainfall data of 57 years (1951-2007). The grid wise rainfall data was provided by IMD. This study examined the drought patterns across space and time using time series SPI. Despite its coarse resolution, SPI has revealed many interesting results on the variability in the occurrence of meteorological drought in India. Decreasing trend in the area under dryness in June and increasing trend in July has many implications on the agriculture management towards minimizing the impact of early season drought on crop planting. Comparison of drought frequency in different decades and ranking years based on drought persistence are extremely useful for understanding historic drought patterns and assessment of future risk. Increased drought frequency in the recent decade observed in the study facilitates better preparedness and coping mechanism. The SPI based drought patterns can be integrated with agricultural and hydrological parameters for quantifying drought risk. The results of the study are also relevant to climate change studies to understand the historic patterns and build future scenarios of drought.

Further research should include the SPI of higher spatial resolution and relating dryness patterns with cropping pattern and crop production. Such an endeavor enables quantification of drought impact in economic terms.

**Acknowledgements**

We express our sincere thanks to Dr. V. Jayaraman, Director, and National Remote Sensing Centre (NRSC) for his constant encouragement and guidance. Thanks are also due to Dr. R.S. Dwivedi, Group Director, Land Resources Group, NRSC, for his suggestions. Comments and suggestions offered by the anonymous referees have been useful to improve the analysis and manuscript of this paper.

Figure 1a. Spatial extent of dryness of June for the years 1965, 1969, 1974

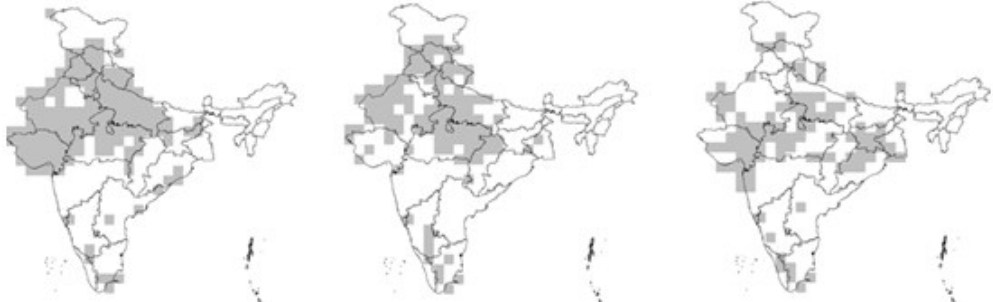

Figure 1b. Spatial extent of dryness of July for the years 2006, 2002, 1987

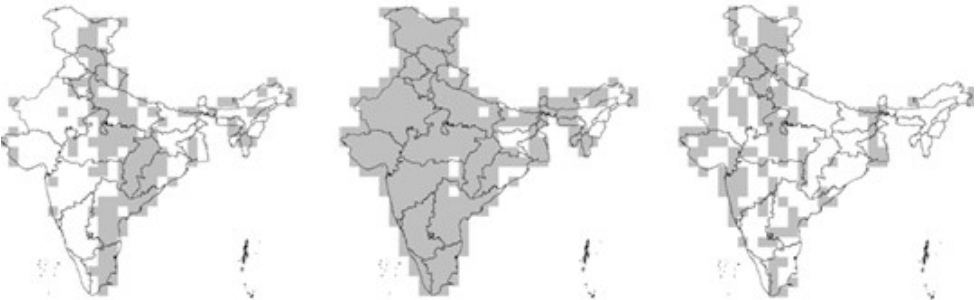

Figure 1c. Spatial extent of dryness of August for the years 1968, 1979, 1987

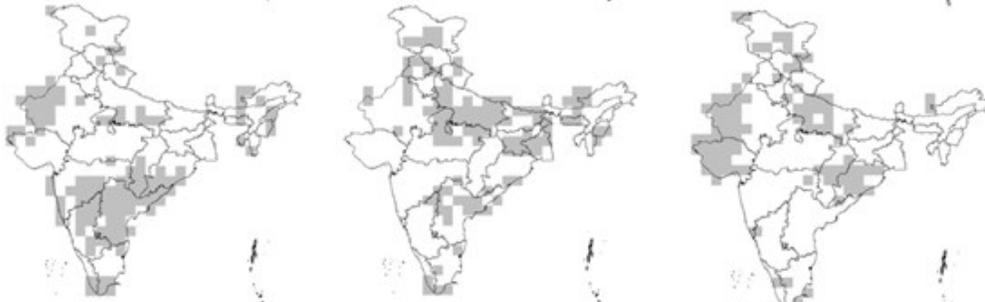

Figure 1d. Spatial extent of dryness of September for the years 1952, 1987, 2001

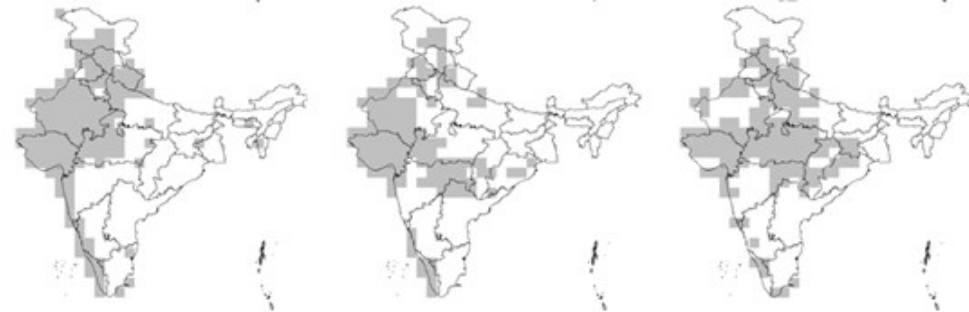

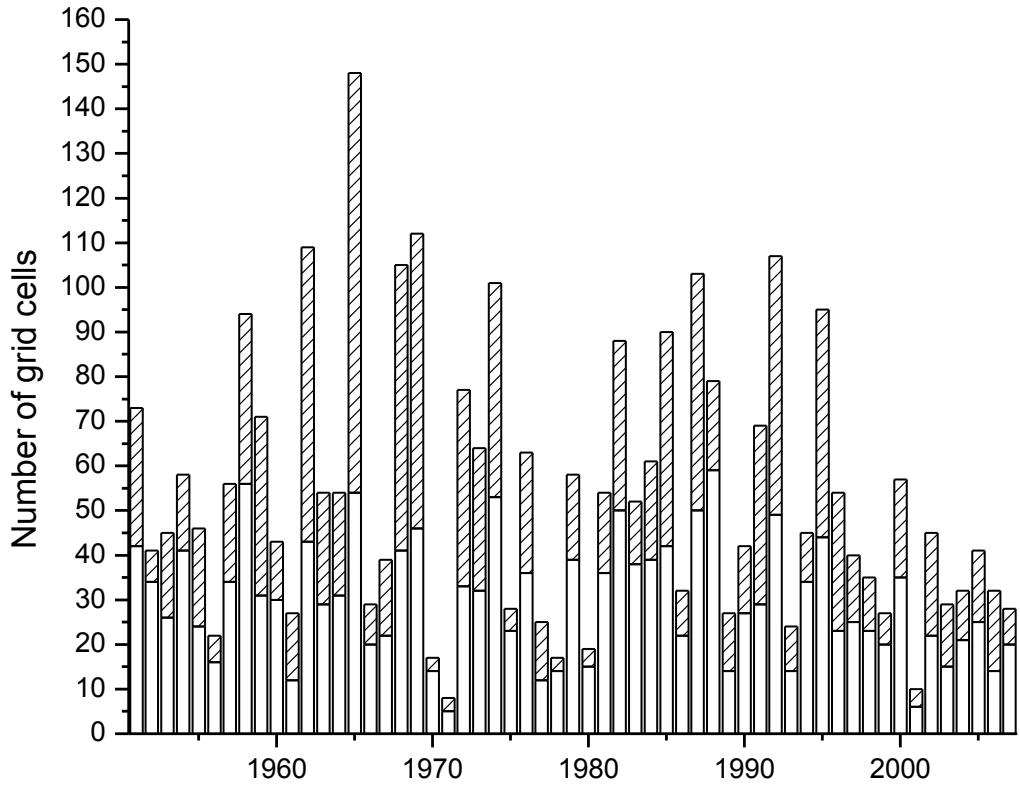

Figure 2a. AUD in June during 1951-2007 (a) blank bars: moderate dryness (b) hatched bars: represent severe-extreme dryness.

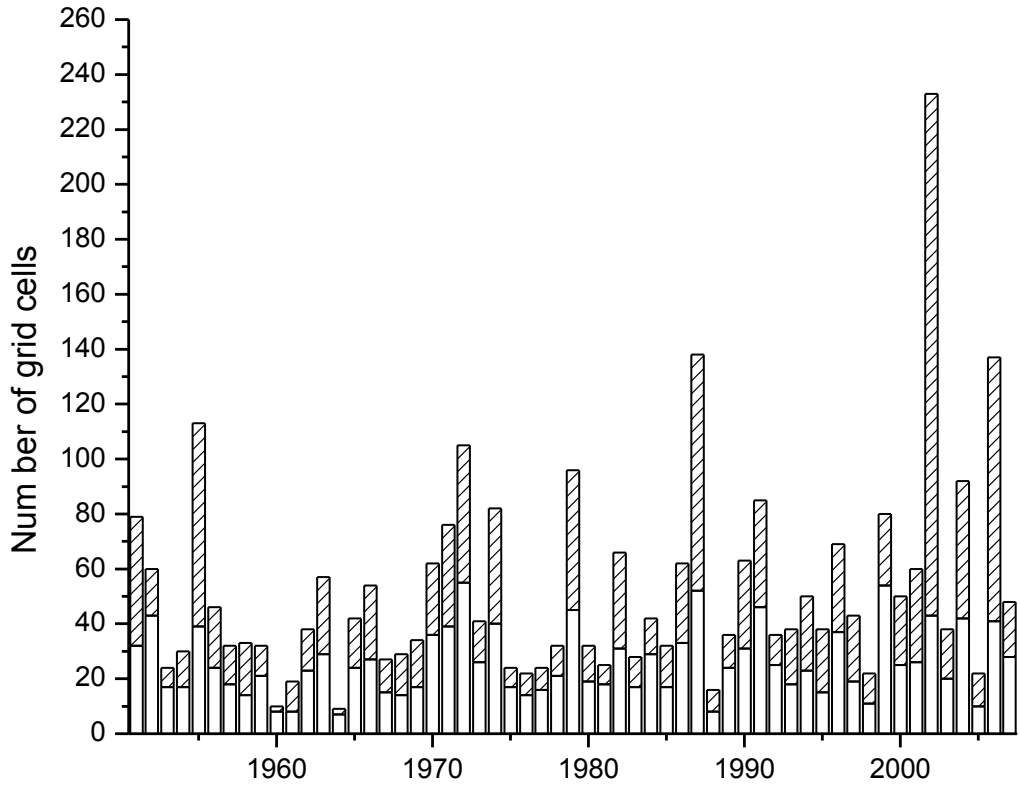

Figure 2b. AUD in July month during 1951-2007 (a) blank bars: moderate dryness (b) hatched bars: severe-extreme dryness.

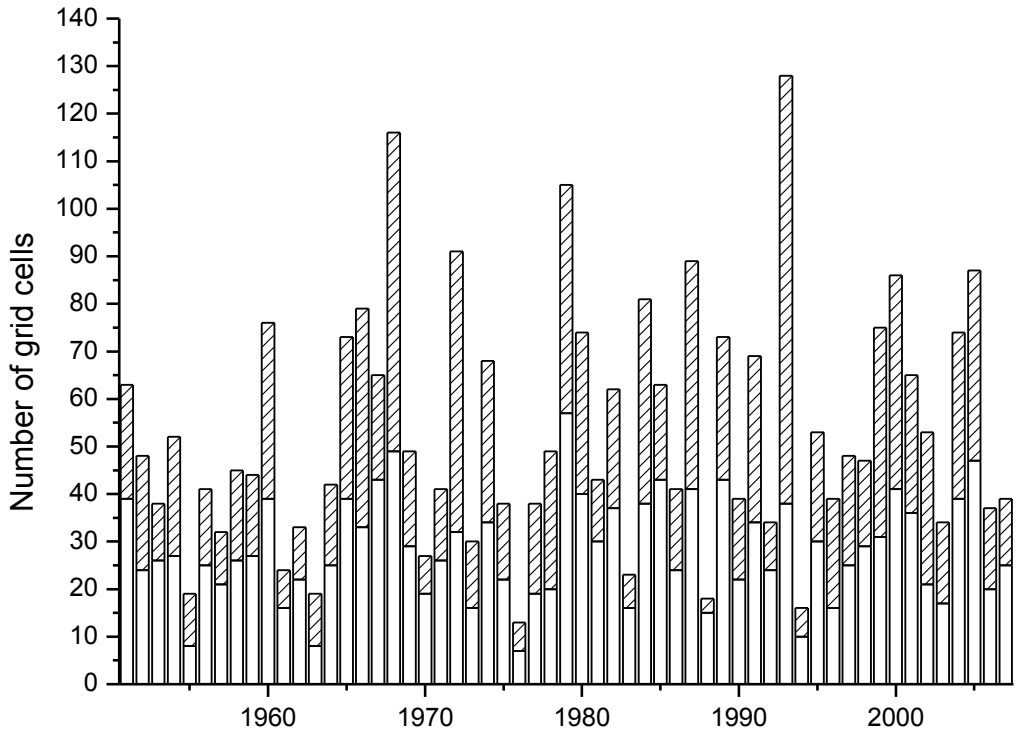

Figure 2c. AUD in August month during 1951-2007 (a) blank bars: moderate dryness (b) hatched bars: represent severe-extreme dryness.

Figure 2d. AUD in September month during 1951-2007 (a) blank bars: moderate dryness (b) hatched bars: severe-extreme dryness.

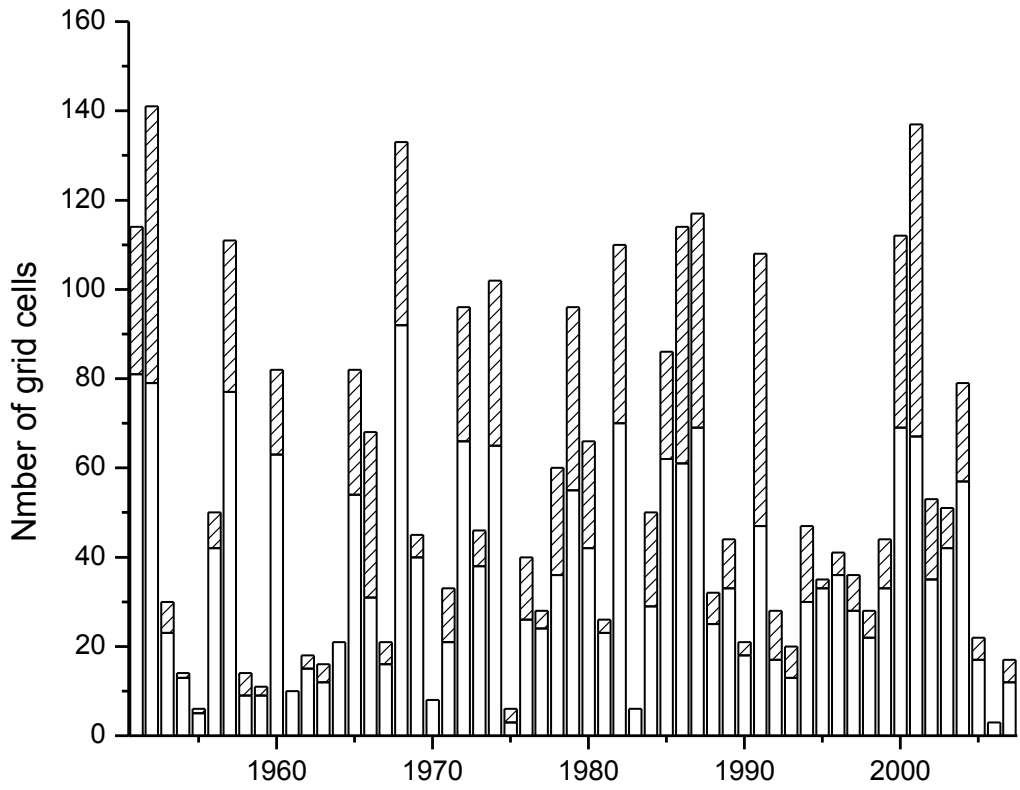

Figure 3. Country wide AUD for June month a) diamond actual AUD b) Line indicates AUD smoothened using LOWESS  c) Line with star indicates increasing trend  d) dotted line indicates decreasing trend

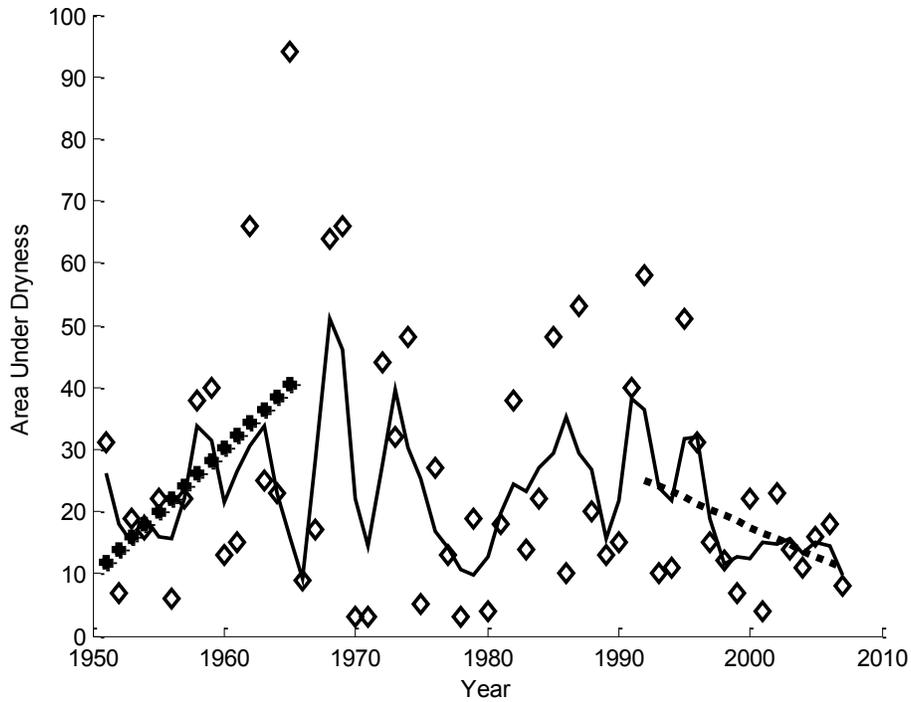

Figure 4a. Drought frequency of 2-3 years for grids from June to September

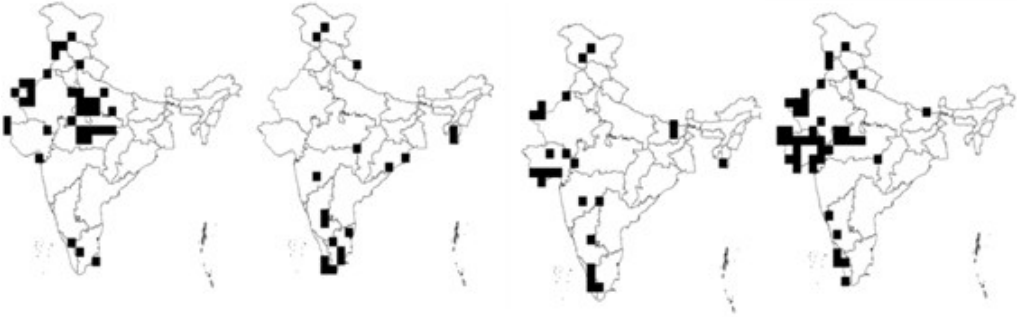

Figure 4b. Drought frequency of 4-5 years for grids from June to September

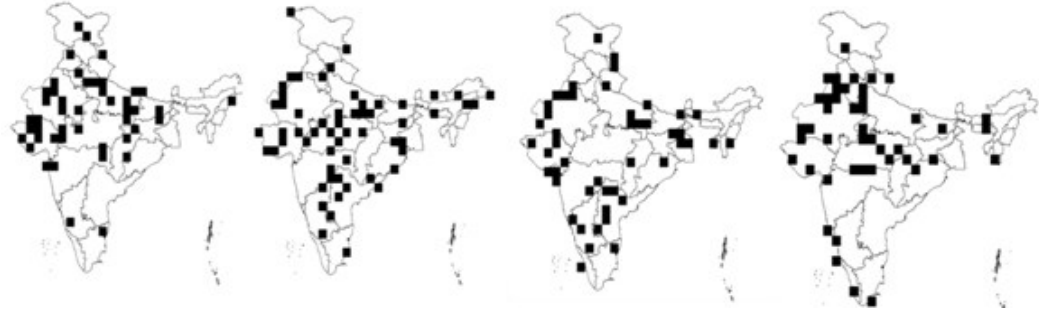

Figure 5a. Decadal drought frequency for the month of June (a) dense lines from left to right 1957 to 1967 (b) sparse lines left to right 1967 to 1977 (c) dense line from right to left 1977 to 1987 (d) sparse lines from right to left 1987 to 1997 (e) dense cross lines 1997 to 2007.

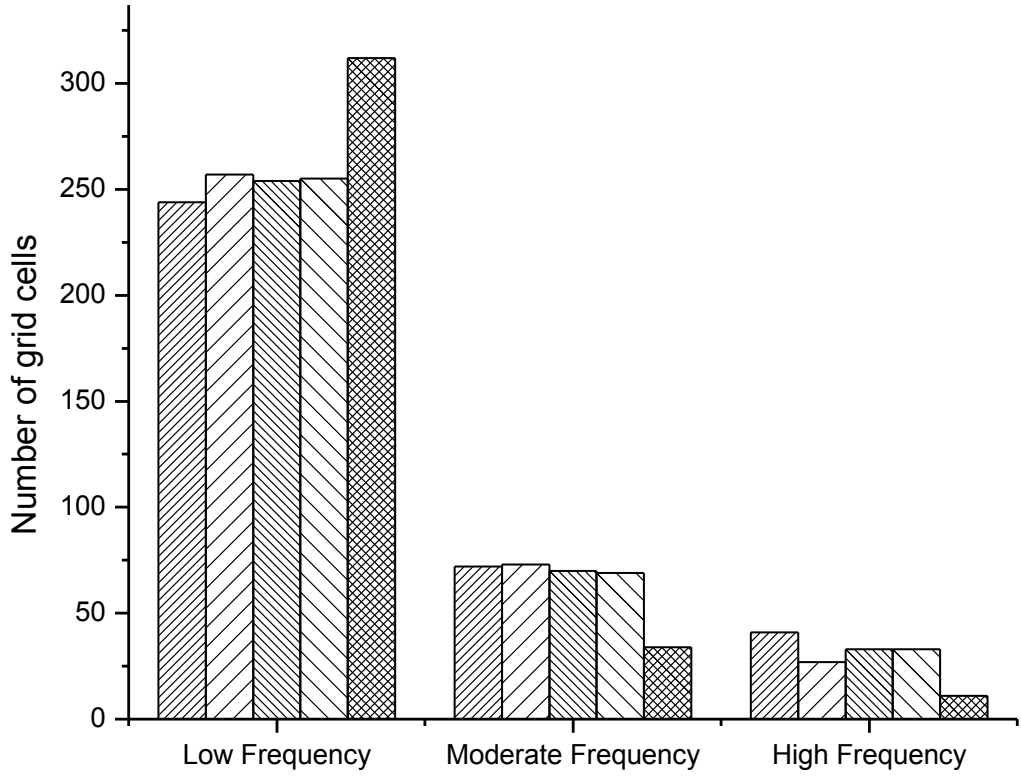

Figure 5b. Decadal drought frequencies for the month of July (a) dense lines from left to right 1967 (b) sparse lines left to right 1977 (c) dense line from right to left 1987 (d) sparse lines from right to left 1997 (e) dense cross lines 2007.

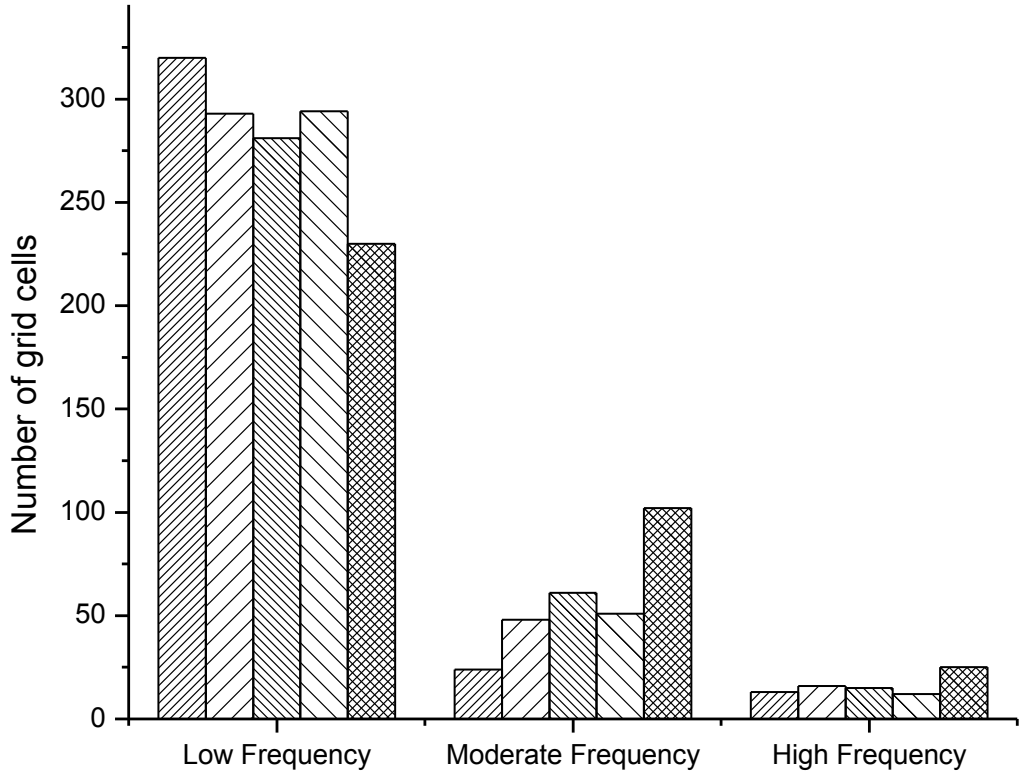

Figure 5c. Decadal drought frequencies in August left to right lines (a) dense lines from left to right 1967 (b) sparse lines left to right 1977 (c) dense line from right to left 1987 (d) sparse lines from right to left 1997 (e) dense cross lines 2007.

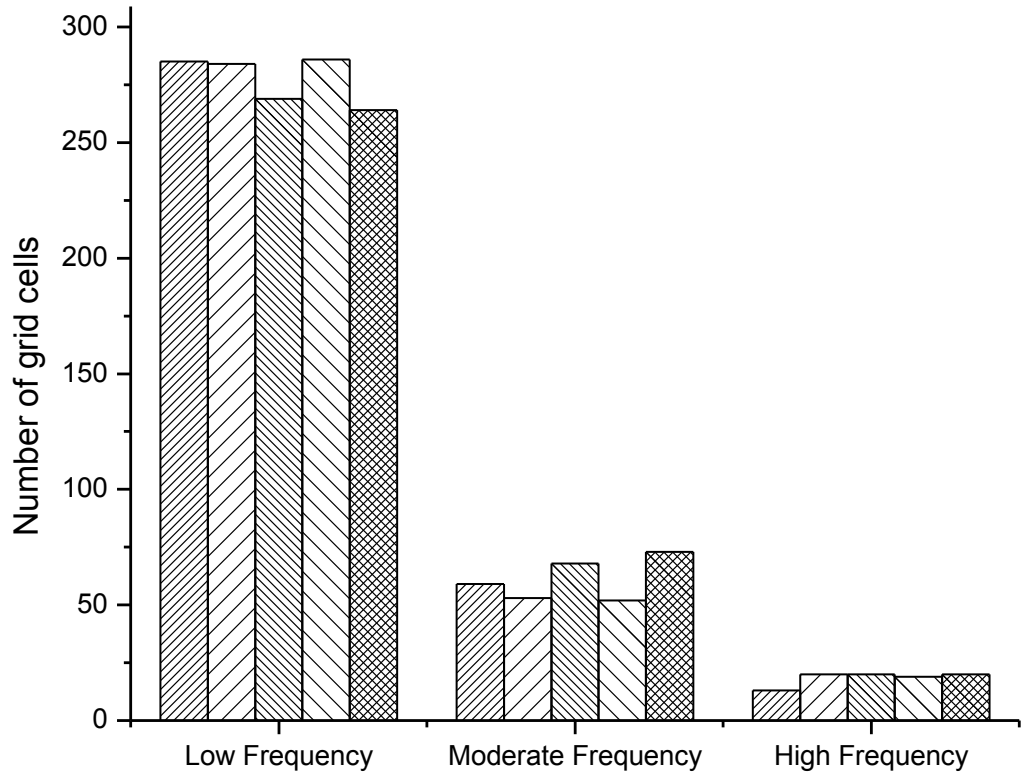

Figure 5d. Decadal drought frequencies in September left to right lines (a) dense lines from left to right 1967 (b) sparse lines left to right 1977 (c) dense line from right to left 1987 (d) sparse lines from right to left 1997 (e) dense cross lines 2007.

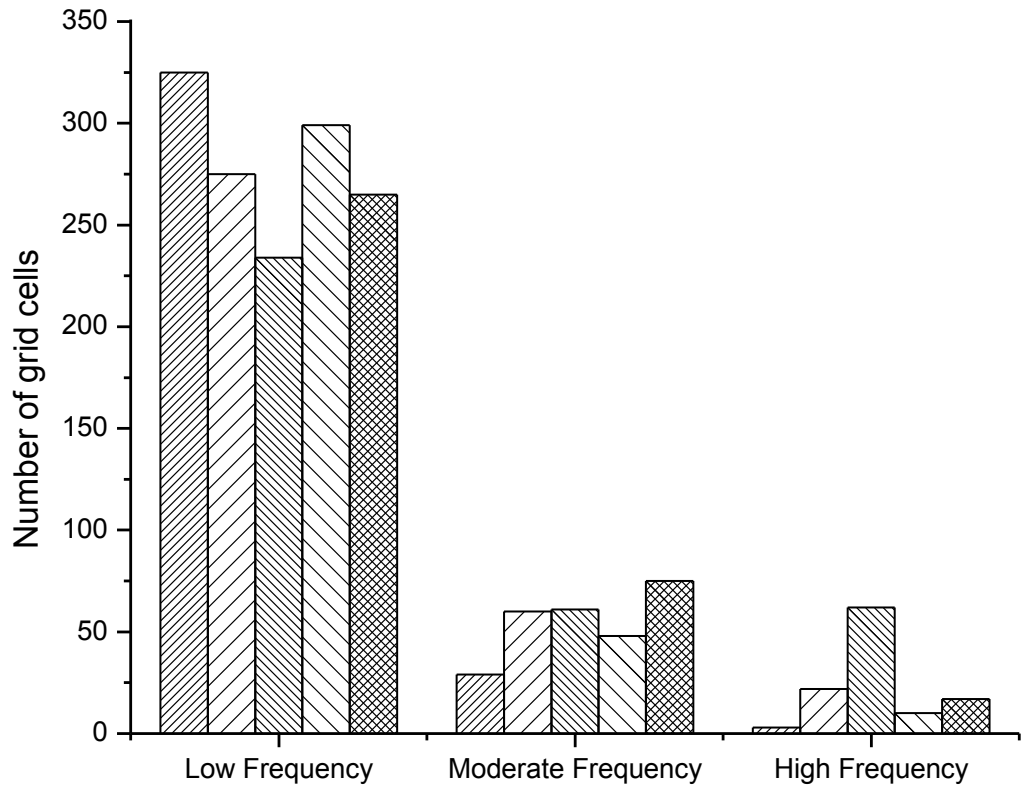

Figure 6. Drought Persistency Score (DPS) of different years (a) blank bars: non-persistent drought (b) crossed bars: persistent drought

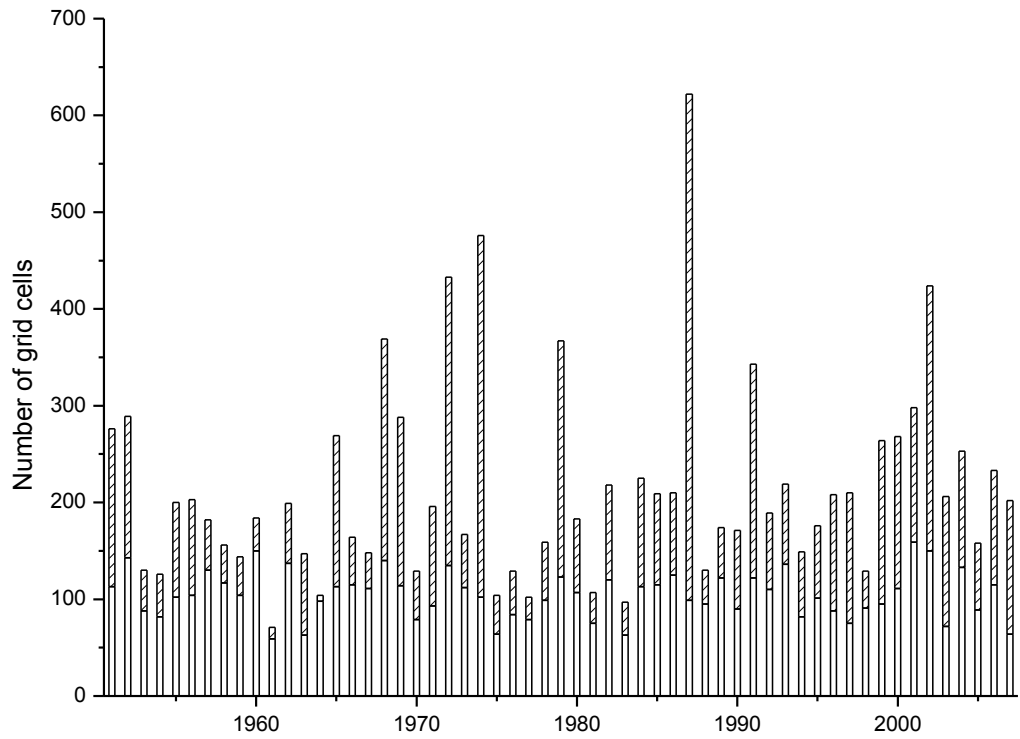

**Table 1 Trend equations with moving averages**

|           | 5-year moving average | 7-year moving average |
|-----------|------------------------|------------------------|
| June      | -0.362 x +64.94, $R^2$ = 0.175 | -0.377 x + 65.55, $R^2$ = 0.230 |
| July      | 0.662x + 34, $R^2$ = 0.333 | 0.68 x + 33.31, $R^2$ = 0.463 |
| August    | 0.312 x + 45.15, $R^2$ = 0.228 | 0.309 x + 45.78, $R^2$ = 0.328 |
| September | 0.308 x +44.57, $R^2$ = 0.099 | 0.336 x + 44.58, $R^2$ = 0.164 |

Table 2 Grouping of years on the basis of drought persistency

| Category | Years in the category |
|----------|------------------------|
| < 100 (very low drought intensity) | 1961, 1964, 1975, 1977, 1981, 1983 |
| 100-200 (Low drought intensity) | 1953, 1954, 1955, 1956, 1957,1958, 1959, 1960, 1962, 1963, 1966, 1967, 1970, 1971, 1973, 1976, 1978, 1980, 1985, 1986, 1988, 1989, 1990,1992, 1994, 1995, 1996, 1997, 1998, 2003, 2005, 2007 |
| 200-300 (Moderate drought intensity) | 1951, 1952, 1965, 1969, 1982, 1984, 1993, 1999, 2000, 2001, 2004, 2006 |
| 300-400 (High drought intensity) | 1968, 1979,1991 |
| 400-500 (Very high drought intensity) | 1972, 1974, 2002 |
| >600 (Extremely high drought intensity) | 1987 |